\documentclass[runningheads]{llncs}
\usepackage[T1]{fontenc}
\usepackage{graphicx,verbatim,multirow,textcomp,booktabs,pifont,array,tikz,nicematrix,makecell}
\usepackage{amssymb,amsfonts}
\usepackage{algorithm,algorithmic}
\usepackage{hyperref}
\usepackage{textcomp}
\usepackage{adjustbox}
\usepackage{xcolor}

\newcolumntype{P}[1]{>{\centering\arraybackslash}p{#1}}
\newcolumntype{M}[1]{>{\centering\arraybackslash}m{#1}}

\newcommand{\cmark}{\ding{51}} 
\begin{document}
\title{ViPSAM: Visual Prompting Medical Image Segmentation Using Segment Anything Model}
\titlerunning{ViPSAM: Visual Prompting Medical Image Segmentation Using SAM}
%

\author{San Lee\inst{1} \and
Nalee Kim\inst{2} \and
Jeong Il Yu\inst{2} \and
Hee Chul Park\inst{2} \and
Boah Kim\inst{3}\thanks{Corresponding author. Email: boah.kim at skku.edu}}

\index{Lee, San}
\index{Kim, Nalee}
\index{Yu, Jeong Il}
\index{Park, Hee Chul}
\index{Kim, Boah}

\authorrunning{S. Lee et al.}

\institute{Department of Artificial Intelligence, Sungkyunkwan University, Republic of Korea \and
Department of Radiation Oncology, Samsung Medical Center,\\ 
Sungkyunkwan University School of Medicine, Republic of Korea \and
Department of MetaBioHealth, Sungkyunkwan University, Republic of Korea
}
  
\maketitle              

\begin{abstract}
In proton therapy planning, respiratory-gated non-contrast CT (NCCT) is commonly used for lesion segmentation; however, accurate delineation remains challenging due to low lesion-to-background contrast. Although learning-based methods have shown strong performance, they often struggle with non-contrast image segmentation. Inspired by clinical practice, where contrast-enhanced MRI is referenced to delineate lesions on NCCT, we propose ViPSAM, a visual prompting framework that leverages complementary cross-modality information. Built upon the Segment Anything Model (SAM), ViPSAM introduces a visual prompt encoder to extract guidance features from contrast-enhanced images and a visual-guided cross-attention module to integrate non-contrast and contrast-enhanced features, thereby enhancing lesion-relevant representations in low-contrast regions. The mask decoder is further adapted in a parameter-efficient manner to utilize visual prompts effectively. We evaluate the proposed method on liver lesion segmentation using NCCT acquired for proton therapy. Experimental results demonstrate that ViPSAM outperforms representative U-Net- and SAM-based methods, indicating that cross-modality visual prompting enables more robust and accurate segmentation in non-contrast images.

\end{abstract}

\section{Introduction}
Accurate lesion delineation is important for proton therapy planning of liver tumors, as it determines target localization and proton beam delivery \cite{cheng2020proton,newhauser2015physics}. In clinical workflows, respiratory-gated non-contrast CT (NCCT) serves as the reference image for treatment planning \cite{cheung2022evaluation}. However, precise segmentation on NCCT is challenging due to respiratory-induced anatomical variations and the inherently low contrast between lesions and surrounding liver regions \cite{iannaccone2005hepatocellular,tsai2018quantitative,chang2017consensus,velec2011effect}. To address this, clinicians often refer to contrast-enhanced MRI acquired in a specific respiratory phase to better identify the lesion boundaries while contouring on NCCT. Nevertheless, the manual process is time-consuming and prone to inter-observer variability \cite{sherer2021metrics,marshall2023interobserver}. These challenges highlight the need for robust automated segmentation methods for low-contrast treatment planning images.

With advances in deep learning, medical image segmentation has been extensively studied using U-Net-based architectures \cite{ronneberger2015u} and their extensions \cite{isensee2021nnu,chen2021transunet,cao2022swin}. Although these models have achieved outstanding performance on various tasks, they typically operate in a single-modality setting and rely solely on intensity-based appearance cues. Accordingly, in non-contrast medical images, the low lesion-to-background contrast reduces feature discriminability, making it difficult for models to segment lesion boundaries with high performance. 

Recently, foundation models trained on large-scale datasets have demonstrated strong transferability across downstream tasks, including image segmentation \cite{kirillov2023segment,zhang2023input,shi2023generalist,cox2024brainsegfounder}. In particular, the Segment Anything Model (SAM) \cite{kirillov2023segment} presents a prompt-based segmentation framework and has shown remarkable generalization in medical imaging applications \cite{ma2024segment,zhang2023customized}. However, when applied to low-contrast images such as NCCT, SAM-based models still struggle to delineate small lesions with indistinct boundaries. Moreover, adapting foundation models to specific medical domains often requires substantial computational cost. 

To address these challenges, we propose a visual prompting medical image segmentation model based on SAM, termed \textit{ViPSAM}, which leverages complementary multi-modal information within a unified framework. Inspired by the clinical practice where radiation oncologists reference contrast-enhanced MRI to delineate lesions on NCCT, our model incorporates cross-modality visual guidance to enhance segmentation in low-contrast clinical scenarios.

Specifically, we introduce a visual prompt encoder to extract soft-tissue contrast cues from contrast-enhanced images, which serve as visual prompts and provide complementary information to non-contrast medical images. These cues are integrated with non-contrast image representations through a visual-guided cross-attention module, enabling the model to better capture lesion-relevant information. Furthermore, we adapt the SAM mask decoder by incorporating Low-Rank Adaptation (LoRA) \cite{hu2022lora} for parameter-efficient fine-tuning. Through this design, segmentation is performed on non-contrast images, while contrast-enhanced images are used solely as visual prompts.

We evaluate ViPSAM for liver lesion segmentation using a proton therapy dataset collected at a tertiary medical center, consisting of NCCT and corresponding contrast-enhanced MRI. Segmentation is performed on NCCT using MRI-derived visual prompts. Experimental results demonstrate that visual prompting significantly improves segmentation performance, surpassing representative U-Net- and SAM-based methods. Our contributions are as follows:

\begin{itemize}
  \item We propose ViPSAM, a SAM-based visual prompting framework leveraging cross-modal cues for lesion segmentation in non-contrast medical images.
  \item We design a visual prompt encoder and a visual-guided cross-attention module that conditions non-contrast features on contrast-enhanced visual prompts, together with LoRA-based parameter-efficient adaptation within SAM.
  \item Experimental results on a liver lesion segmentation dataset for proton therapy with patient-matched NCCT and contrast-enhanced MRI demonstrate superior performance over representative U-Net- and SAM-based methods.
\end{itemize}

\begin{figure}[t]
  \centering
  \includegraphics[width=\linewidth]{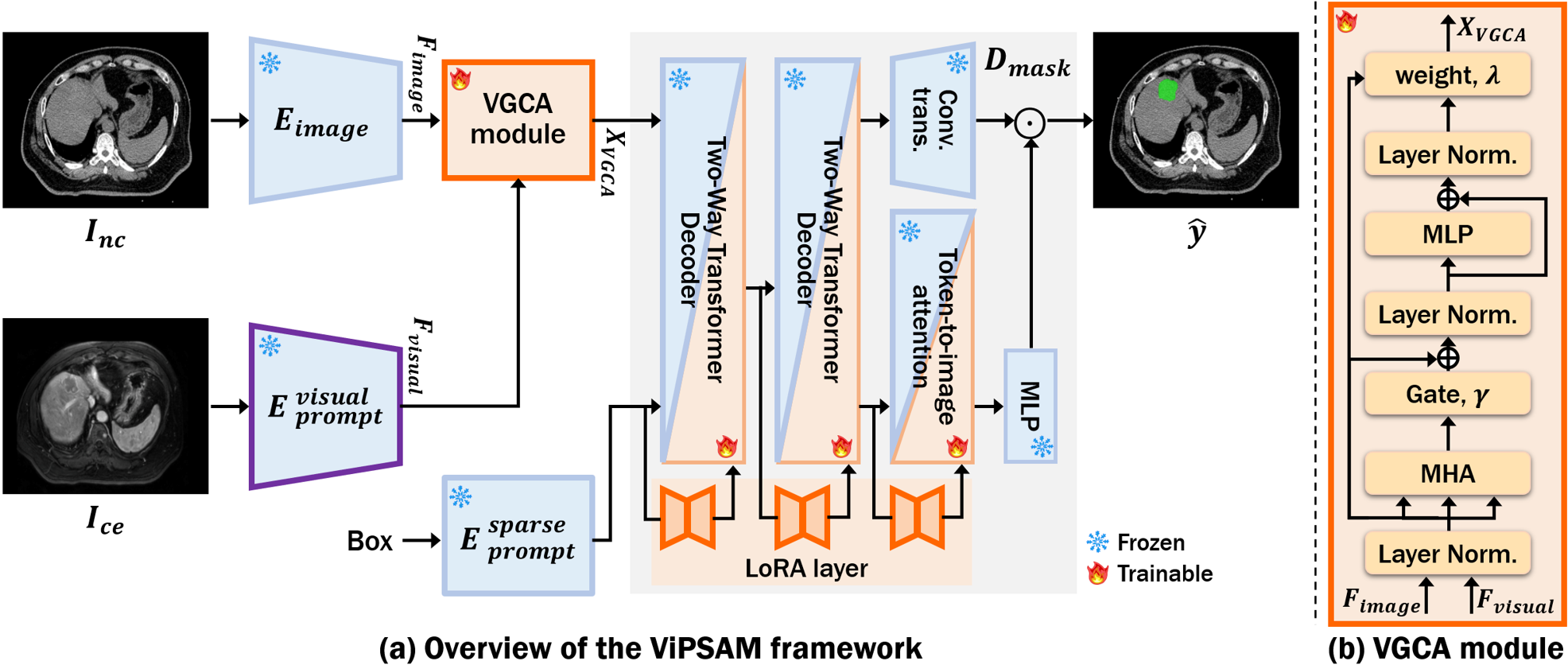}
  \caption{(a) Overview of ViPSAM. Lesion segmentation $\hat{y}$ on non-contrast images $I_{nc}$ is predicted by leveraging complementary information from contrast-enhanced images $I_{ce}$ via a visual prompt encoder, a visual-guided cross-attention (VGCA) module, and LoRA-based adaptation in the mask decoder. (b) Architecture of the VGCA module.
}
  \label{fig:arch}
\end{figure}

\section{Proposed Method}
An overview of ViPSAM, our SAM-based framework for visual-prompting image segmentation, is shown in Fig.~\ref{fig:arch}. ViPSAM extends the Segment Anything Model (SAM) by incorporating (i) a visual prompt encoder that extracts soft-tissue contrast cues from cross-modality contrast-enhanced images, (ii) a visual-guided cross-attention module that conditions non-contrast image features on the extracted visual prompt representations, and (iii) Low-Rank Adaptation (LoRA) within the decoder for parameter-efficient fine-tuning. We first provide a brief overview of SAM, followed by detailed descriptions of each component.

\subsubsection*{Segment Anything Model (SAM)} SAM \cite{kirillov2023segment} consists of an image encoder $E_{image}$, a prompt encoder $E_{prompt}$, and a mask decoder $D_{mask}$. The image encoder, implemented as a Vision Transformer (ViT), takes an input image $I$ and extracts image embedding $F_{image}=E_{image}(I)$. The prompt encoder embeds a prompt $P$ (e.g., points, boxes, or masks) into prompt tokens $E_{prompt}(P)$. The mask decoder then takes the image embedding $F_{image}$ and the prompt tokens $E_{prompt}(P)$ to predict a segmentation mask $\hat{y}=D_{mask}(F_{image}, E_{prompt}(P))$ via Two-Way Transformer blocks that perform bidirectional cross-attention between the prompt tokens and image embedding. In our framework, we adopt SAM with a ViT-B image encoder and keep its pretrained parameters frozen during training, using box prompts as input.

\subsection{Visual Prompt Encoder}
\label{sec:visual prompt encoder}
To mitigate low lesion-to-background contrast in non-contrast images, we introduce a visual prompt encoder $E_{prompt}^{visual}$, where \textit{visual prompt} refers to image-based guidance from contrast-enhanced images that provide clearer lesion contrast, in contrast to sparse prompts (e.g., boxes). 
Notably, $E_{prompt}^{visual}$ uses the same ViT-B architecture and SAM-pretrained weights as the image encoder $E_{image}$, with all weights frozen during training.

Given a non-contrast image $I_{nc}$ and a contrast-enhanced image $I_{ce}$, each image is replicated to three channels to match the SAM input format, yielding $I_{nc}, I_{ce} \in \mathbb{R}^{3 \times H \times W}$,
where $H$ and $W$ represent the height and width of the input image, respectively. When the non-contrast image is fed into $E_{image}$ to extract: 
\begin{align}
F_{image} = E_{image}(I_{nc})\in\mathbb{R}^{c\times h\times w}, 
\end{align}
where $c$ and $h\times w$ denote the channel dimension and spatial resolution of the feature map, respectively, the contrast-enhanced image is encoded by $E_{prompt}^{visual}$ to obtain visual prompt features:
\begin{align}
F_{visual} = E_{prompt}^{visual}(I_{ce})\in\mathbb{R}^{c\times h\times w}.
\end{align}
Since both encoders have an identical architecture, the resulting feature maps can interact directly in the subsequent visual-guided cross-attention module.

\subsection{Visual-Guided Cross-Attention Module}
To obtain visual-prompt-guided representations with the non-contrast image serving as the spatial reference, we design a visual-guided cross-attention (VGCA) module (Fig.~\ref{fig:arch}(b)) that conditions non-contrast image features $F_{image}$ on visual prompt features $F_{visual}$:
\begin{align}
    X_{VGCA} = \mathrm{VGCA}(F_{image}, F_{visual}).
\end{align}

Specifically, in the VGCA module, $F_{image}$ and $F_{visual}$ are first flattened and normalized by layer normalization (LN) \cite{ba2016layer}:
\begin{align}
X_{image} = \mathrm{LN}(\mathrm{Flat}(F_{image})), \quad
X_{visual} = \mathrm{LN}(\mathrm{Flat}(F_{visual})),
\end{align}
where $\mathrm{Flat}(\cdot)$ reshapes $\mathbb{R}^{c \times h \times w}$ to $\mathbb{R}^{n \times c}$ with $n= h\cdot w$. 
Then, multi-head cross-attention (MHA) is computed as:
\begin{align}
A = \mathrm{MHA}(Q=X_{image}, K=X_{visual}, V=X_{visual}),
\end{align}
where Q, K, and V denote the query, key, and value, respectively. Since segmentation is performed on non-contrast images, $X_{\mathrm{image}}$ is used as the query, while $X_{\mathrm{visual}}$ serves as the key and value to provide complementary lesion information from contrast-enhanced images.

To regulate the influence of visual prompts, a learnable gating scalar $\gamma$ is introduced through a gated residual update:
\begin{align}
X_\gamma &= \mathrm{LN}(X_{image} + \gamma A), \\
\tilde{X} &= \mathrm{LN}(X_\gamma + \mathrm{MLP}(X_\gamma)),
\end{align}
where MLP denotes a feed-forward network with a Linear–GeLU–Linear structure, and $\tilde{X}$ is the cross-attended representation.
The visual-prompt-conditioned image embedding $X_{VGCA}$, used in the mask decoder, is then computed as:
\begin{align}
X_{VGCA} = \lambda X_{image} + (1-\lambda)\tilde{X},
\end{align}
where $\lambda$ is a learnable weight scalar. 
Unlike standard cross-attention, our formulation with learnable gating and weighting preserves the non-contrast image as the primary spatial reference while enabling cross-modal visual guidance to modulate feature representations.

\begin{figure}[t]
  \centering
  \includegraphics[width=\linewidth]{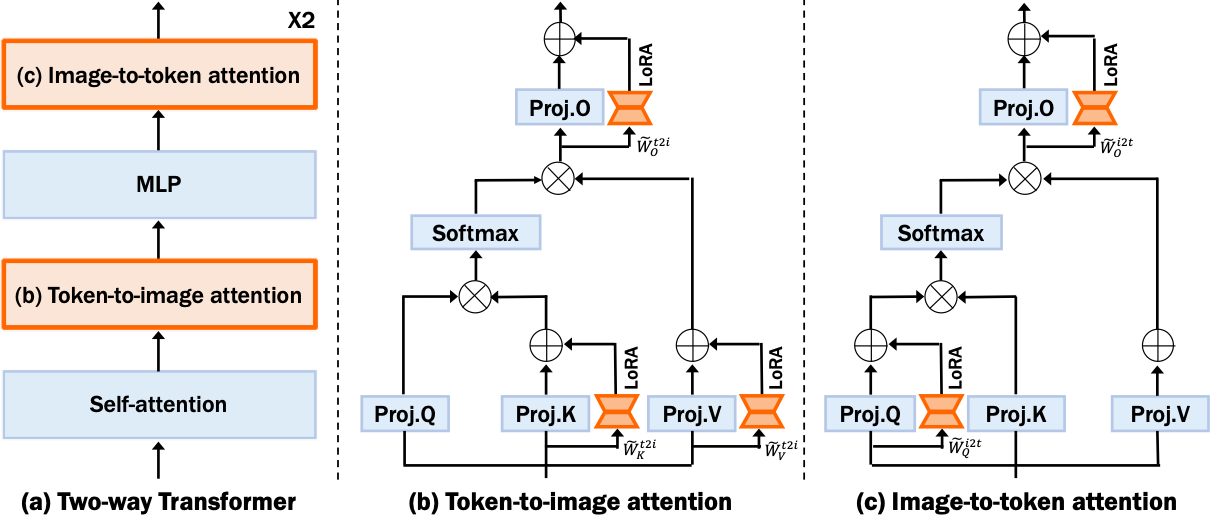}
  \caption{LoRA placement in the SAM mask decoder. (a) Two-Way Transformer block in the mask decoder. (b,c) LoRA is applied to selected projections in cross-attention layers: token-to-image $(K, V, O)$ and image-to-token $(Q, O)$.}
  \label{fig:decoder_lora}
\end{figure}

\subsection{LoRA in Mask Decoder}
\label{sec:lora_decoder}
In our framework, the mask decoder $D_{mask}$ consists of stacked Two-Way Transformer blocks that alternate token-to-image and image-to-token cross-attention (Fig.~\ref{fig:decoder_lora}(a)). This takes sparse prompt tokens generated by the sparse prompt encoder $E_{prompt}^{sparse}$ from a box prompt $P$ which provides localization cues, together with image embeddings produced by the VGCA module, and predicts the segmentation mask $\hat{y}$ by:
\begin{align}
    \hat{y} = D_{mask}(X_{VGCA}, E_{prompt}^{sparse}(P)).
\end{align}

Here, to effectively adapt the mask decoder to visual-prompt-conditioned image embeddings $X_{VGCA}$ in a parameter-efficient manner, we incorporate Low-Rank Adaptation (LoRA) \cite{hu2022lora} into the decoder by updating frozen linear projections with learnable low-rank residuals. Details are described below.

Let $X \in \mathbb{R}^{n \times c_{\text{in}}}$ denote an image embedding and $W \in \mathbb{R}^{c_{\text{out}} \times c_{\text{in}}}$ be a frozen linear projection producing $\hat{X} = WX$. In the LoRA layer, the adapted weight is given by:
\begin{align}
\label{eq:lora_basic}
\tilde{W} = W + BA,
\end{align}
where $A \in \mathbb{R}^{r \times c_{\text{in}}}$ and $B \in \mathbb{R}^{c_{\text{out}} \times r}$ are learnable linear layer with rank $r \ll \min(c_{\text{in}}, c_{\text{out}})$. During training, $W$ remains fixed and only $(A, B)$ are updated, enabling parameter-efficient adaptation with minimal additional parameters.

Following this strategy, we insert LoRA into selected cross-attention projections of the mask decoder according to the role of image embeddings in each pathway.
Specifically, in token-to-image (t2i) attention (Fig.~\ref{fig:decoder_lora}(b)), sparse prompt tokens serve as queries ($Q$) to retrieve spatially relevant image cues for target localization, while image embeddings act as keys ($K$) and values ($V$). Since image embeddings are projected into $K$ and $V$, LoRA is applied to the key, value, and output projection matrices, i.e., $\tilde{W}_{K}^{t2i}$, $\tilde{W}_{V}^{t2i}$, and $\tilde{W}_{O}^{t2i}$.
In contrast, in image-to-token (i2t) attention (Fig.~\ref{fig:decoder_lora}(c)), image embeddings serve as queries to attend back to sparse prompt tokens for feature refinement, while sparse prompt tokens act as keys and values. Accordingly, LoRA is applied to the query and output projection matrices, i.e., $\tilde{W}_{Q}^{i2t}$ and $\tilde{W}_{O}^{i2t}$.

Together, the t2i and i2t attention layers enable bidirectional interaction between sparse prompts and dense image features, while selective LoRA placement facilitates efficient adaptation of the mask decoder to visual-prompt-conditioned representations without full fine-tuning, resulting in accurate lesion segmentation on non-contrast images.

\begin{figure}[t!]
  \centering
  \includegraphics[width=\linewidth]{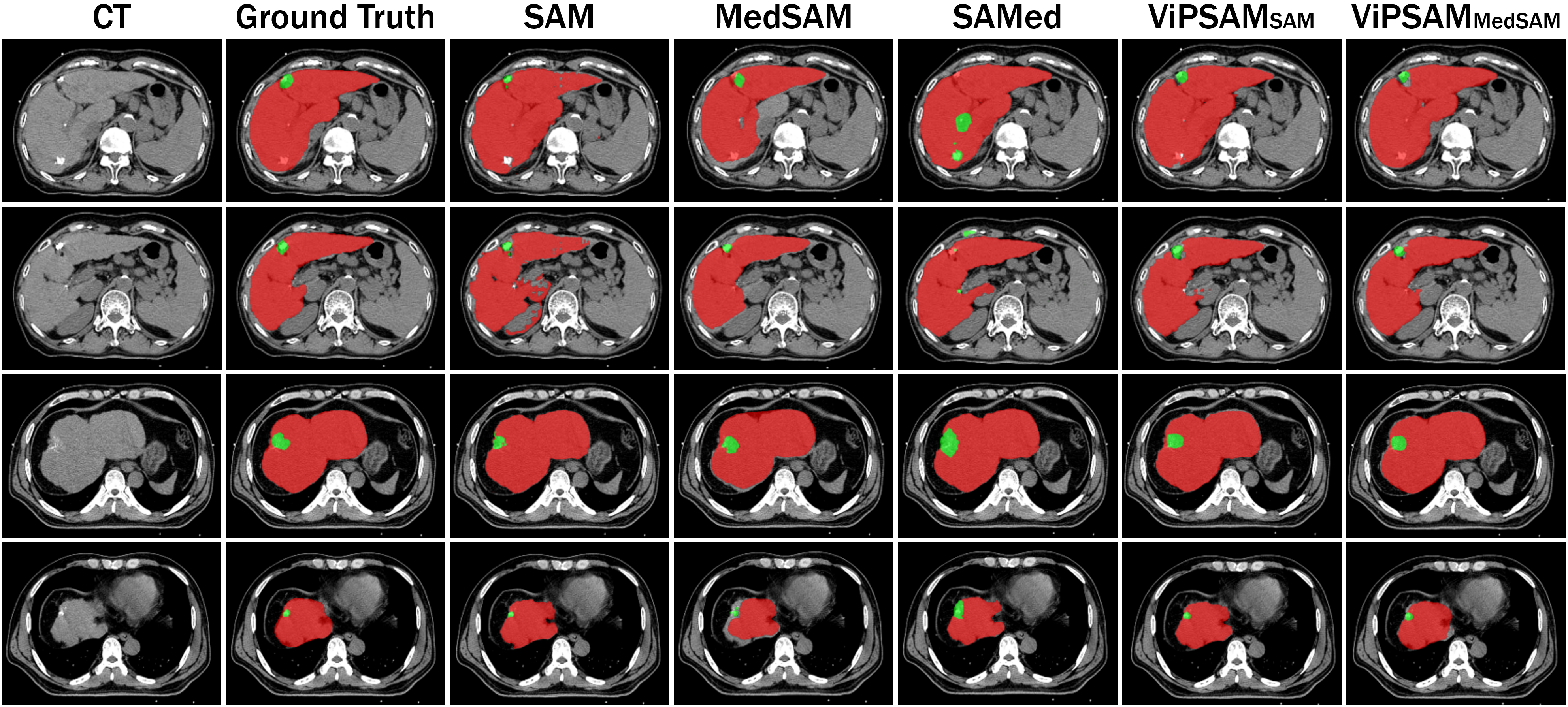}
  \caption{Qualitative comparison results of liver lesion segmentation on non-contrast CT. Red denotes the liver and green denotes the liver lesion.}
  \label{fig:Qualitative_comparison}
\end{figure}

\section{Experiments}

\subsubsection*{Dataset and Metric} 
To evaluate the proposed cross-modality visual prompting framework, we used a liver lesion dataset for proton therapy planning collected at Samsung Medical Center. The dataset comprises 73 cases, each including a mid-respiratory phase NCCT (T = 50\%), a T1-weighted fat-suppressed contrast-enhanced MRI at the same respiratory phase, and expert-annotated liver and lesion masks defined on the NCCT.
MRI scans were rigidly registered to the NCCT, and all scans were resampled to a voxel spacing of $0.6597 \times 0.6597 \times 5$ mm. NCCT scans were clipped to [-100, 200] HU. For both CT and MRI, each 2D axial slice was resized to $1024 \times 1024$ and normalized to [0, 1] using min-max scaling. The dataset was split into 60, 6, and 7 cases for training, validation, and testing, respectively. In total, 2,251, 213, and 245 2D axial slices containing liver lesions were used for each split.
For quantitative evaluation, segmentation performance was assessed using the Dice similarity coefficient (DSC), Intersection over Union (IoU), and the 95th percentile Hausdorff distance (HD95).

\begin{table}[t!]
\caption{Quantitative comparison results of the average Dice, IoU, HD95 and the number of trainable parameters. Bold indicates the best performance for each metric.}
\label{tab:smc_liver}
\centering
\resizebox{\linewidth}{!}{
    \begin{tabular}{cccccccc}
\toprule
\multirow{2}{*}{Methods} &
\multicolumn{3}{c}{Liver} &
\multicolumn{3}{c}{Lesion} &
\multirow{2}{*}{\thead[c]{Params\\(M) $\downarrow$}} \\
\cmidrule(lr){2-4}\cmidrule(lr){5-7}
& Dice $\uparrow$ & IoU $\uparrow$ & HD95(mm) $\downarrow$
& Dice $\uparrow$ & IoU $\uparrow$ & HD95(mm) $\downarrow$
& \\
\midrule
U-Net      & 0.875$\pm$0.162 & 0.805$\pm$0.187 & 27.62$\pm$31.13 & 0.236$\pm$0.298 & 0.174$\pm$0.240 & 60.42$\pm$31.35 & 4.32   \\
nnU-Net    & 0.856$\pm$0.238 & 0.799$\pm$0.244 & 28.72$\pm$41.09 & 0.267$\pm$0.339 & 0.209$\pm$0.281 & 43.23$\pm$28.12 & 59.18  \\
TransUNet & 0.861$\pm$0.198 & 0.793$\pm$0.219 & 22.29$\pm$26.83 & 0.169$\pm$0.189 & 0.106$\pm$0.132 & 77.92$\pm$46.17 & 105.28 \\
Swin-Unet & 0.880$\pm$0.155 & 0.809$\pm$0.178 & 24.38$\pm$26.85 & 0.162$\pm$0.189 & 0.101$\pm$0.128 & 56.25$\pm$25.63 & 27.17  \\
SAM       & 0.890$\pm$0.111 & 0.815$\pm$0.138 & 18.80$\pm$13.27 & 0.655$\pm$0.208 & 0.520$\pm$0.215 & 13.11$\pm$10.02 & 93.74  \\
MedSAM    & 0.892$\pm$0.105 & 0.817$\pm$0.136 & 13.59$\pm$12.29 & 0.709$\pm$0.180 & 0.577$\pm$0.206 & 8.39$\pm$4.76   & 93.74  \\
SAMed     & 0.901$\pm$0.128 & 0.838$\pm$0.155 & 18.88$\pm$27.22 & 0.341$\pm$0.328 & 0.259$\pm$0.270 & 48.59$\pm$28.12 & 3.93   \\
\midrule
\textbf{ViPSAM$_{\mathrm{SAM}}$} &
0.925$\pm$0.088 & 0.869$\pm$0.114 & 14.58$\pm$17.68 &
\textbf{0.852$\pm$0.114} & \textbf{0.757$\pm$0.150} & \textbf{5.15$\pm$3.70} &
0.83 \\
\textbf{ViPSAM$_{\mathrm{MedSAM}}$} &
\textbf{0.937$\pm$0.070} & \textbf{0.887$\pm$0.094} & \textbf{8.84$\pm$9.49} &
0.810$\pm$0.151 & 0.704$\pm$0.188 & 5.18$\pm$2.18 &
0.83 \\
\bottomrule
\end{tabular}%
}
\end{table}

\subsubsection*{Implementation Details} 
We implemented the proposed ViPSAM using either SAM \cite{kirillov2023segment} (ViPSAM$_\mathrm{SAM}$) or MedSAM \cite{ma2024segment} (ViPSAM$_\mathrm{MedSAM}$) as the backbone, each initialized with pretrained weights. For the VGCA module, the feature channel dimension was set to $C = 256$ and the number of attention heads was set to 8. We initialize the gating scalar $\gamma$ to 0.3, the weight $\lambda$ to 1.0, and the LoRA rank $r$ to 8. Box prompts were generated from manual masks with random perturbations ($\leq$20 px) during training and fixed during inference.
The model was trained for 12 epochs using an equally weighted sum of Dice and binary cross-entropy losses and optimized with AdamW with a learning rate of $1 \times 10^{-4}$ and a weight decay of $1 \times 10^{-4}$. The checkpoint from the final epoch was used for testing. All experiments were conducted in PyTorch 2.5.1 on an NVIDIA RTX A6000 GPU. Memory usage was approximately 14.3GB during training and 4.8GB during inference. The source code is publicly available at \href{https://github.com/torchViPSAM/ViPSAM}{https://github.com/torchViPSAM/ViPSAM}.

\subsubsection*{Experimental Results} 
We compared ViPSAM with representative segmentation models, including U-Net-based methods (U-Net \cite{ronneberger2015u}, nnU-Net \cite{isensee2021nnu}, TransUNet \cite{chen2021transunet}, and Swin-Unet \cite{cao2022swin}) and SAM-based methods using box prompts (SAM \cite{kirillov2023segment}, MedSAM \cite{ma2024segment}, and SAMed \cite{zhang2023customized}). Baseline models were trained using only NCCT blue with the default settings provided in the original papers, whereas ViPSAM additionally leveraged contrast-enhanced MRI as visual prompts. All methods were evaluated on NCCT for liver and liver lesion segmentation.

Fig.~\ref{fig:Qualitative_comparison} visualizes qualitative comparisons on NCCT. ViPSAM produces more accurate segmentation masks than the existing SAM-based models, even for small lesions with indistinct boundaries. As reported in Table~\ref{tab:smc_liver}, the proposed ViPSAM consistently outperformed both U-Net- and SAM-based methods across all evaluation metrics. 
Specifically, ViPSAM$_{\mathrm{MedSAM}}$ achieved the best liver segmentation performance with a Dice score of 0.937 and an HD95 of 8.84 mm, while ViPSAM$_{\mathrm{SAM}}$ achieved the best lesion segmentation performance with a Dice score of 0.852 and an HD95 of 5.15 mm. 
In contrast, U-Net-based models showed poor lesion segmentation performance under low-contrast NCCT conditions (Dice < 0.267), and SAM-based methods exhibited imprecise boundary delineation for lesions despite box-prompt guidance. Also, our method required only 0.83M trainable parameters, highlighting superior parameter efficiency.

Moreover, we assessed statistical significance using paired t-tests and Wilcoxon signed-rank tests. Both ViPSAM$_{\mathrm{SAM}}$ and ViPSAM$_{\mathrm{MedSAM}}$ significantly outperformed their respective baselines (SAM and MedSAM) across all metrics $(p < 0.05)$ under both tests. These results support the effectiveness of our model leveraging MRI-derived visual prompting for liver lesion segmentation on NCCT.

\begin{table}[t!]
\caption{Ablation study of ViPSAM on prompt design and trainable components.}
\label{tab:ablation_smc_liver}
\centering
\resizebox{\linewidth}{!}{
    \begin{tabular}{ccccccccccc}
\toprule
\multirow{2}{*}{Config.} & \multicolumn{2}{c}{Prompt} & \multicolumn{2}{c}{Layer} &
\multicolumn{3}{c}{Liver} & \multicolumn{3}{c}{Lesion} \\
\cmidrule(lr){2-3}\cmidrule(lr){4-5}\cmidrule(lr){6-8}\cmidrule(lr){9-11}
& Visual & Sparse & CA & LoRA &
Dice $\uparrow$ & IoU $\uparrow$ & HD95(mm) $\downarrow$ &
Dice $\uparrow$ & IoU $\uparrow$ & HD95(mm) $\downarrow$ \\
\midrule
(a) & & \cmark &  & \cmark &
0.909$\pm$0.091 & 0.842$\pm$0.119 & 12.04$\pm$11.02 &
0.801$\pm$0.173 & 0.695$\pm$0.187 & 5.84$\pm$3.14 \\
(b) & \cmark &  & \cmark & \cmark &
0.809$\pm$0.236 & 0.728$\pm$0.247 & 51.85$\pm$44.99 &
0.155$\pm$0.176 & 0.095$\pm$0.116 & 105.91$\pm$43.41 \\
(c) &\cmark & \cmark & \cmark &  &
0.932$\pm$0.081 & 0.881$\pm$0.108 & 9.62$\pm$10.68 &
0.808$\pm$0.136 & 0.669$\pm$0.181 & 5.04$\pm$1.87 \\
Ours & \cmark & \cmark & \cmark & \cmark &
0.937$\pm$0.070 & 0.887$\pm$0.094 & 8.84$\pm$9.49 &
0.810$\pm$0.151 & 0.704$\pm$0.188 & 5.18$\pm$2.18 \\
\bottomrule
\end{tabular}%
}
\end{table}

\subsubsection*{Ablation Study}
To evaluate each component in ViPSAM, we conducted an ablation study with three configurations: (a) ViPSAM without visual prompting, (b) ViPSAM without sparse box prompts, and 
(c) ViPSAM without LoRA-based adaptation. All configurations were trained under the same settings for fair comparison.

Table~\ref{tab:ablation_smc_liver} shows quantitative results for liver and lesion segmentation across the ablated models. Without visual prompting (a), performance decreased to a lesion Dice of 0.801 and an IoU of 0.695, compared to the full ViPSAM (Dice=0.810, IoU=0.704), highlighting the importance of incorporating cross-modal contrast information. Moreover, removing sparse box prompts (b) led to further performance degradation, suggesting their role in stabilizing lesion localization. In addition, without LoRA-based adaptation (c), the model showed reduced performance with a Dice of 0.808 and an IoU of 0.669. 
These results indicate that visual prompting with contrast-enhanced images, VGCA-based cross-modal interaction, and LoRA-based decoder adaptation contribute synergistically to improved liver lesion segmentation on NCCT.

\section{Conclusion}
In this work, we propose ViPSAM, a SAM-based visual prompting framework for non-contrast medical image segmentation. By leveraging complementary information from contrast-enhanced images via a visual prompt encoder, visual-guided cross-attention, and LoRA-based parameter-efficient decoder adaptation, ViPSAM enables precise segmentation with minimal additional learnable parameters. Experimental results on a liver lesion proton therapy planning dataset demonstrate the effectiveness of the proposed framework in alleviating boundary ambiguity in non-contrast imaging, highlighting its potential to enhance segmentation reliability in clinical workflows.


\begin{credits}
\subsubsection{\ackname} This work was supported by the National Research Foundation of Korea(NRF) grant funded by the Korea government(MSIT) (RS-2026-25481239), AI Graduate School Support Program (Sungkyunkwan University) (RS-2019-II190421), and SKKU Academic Research Support Program, Sungkyunkwan University, 2025.

\subsubsection{\discintname}
The authors have no competing interests to declare that are
relevant to the content of this article.
\end{credits}

%
%
%
\bibliographystyle{splncs04}
\bibliography{refs}
%




\end{document}